\documentclass{iaus}
\usepackage{graphicx}

\title[Pulsation-Initiated Mass-Loss in LBVs] 
{Pulsation-Initiated Mass Loss in Luminous Blue Variables: A Parameter Study}

\author[Andrew J. Onifer and Joyce A. Guzik]   
{Andrew J. Onifer$^1$
 \and Joyce A. Guzik$^2$}

\affiliation{$^1$University of Florida, \\Department of Physics, \\PO Box 118440, \\
Gainesville, FL 32611 USA\\
email: {\tt onifer@phys.ufl.edu} \\
[\affilskip]
$^2$Los Alamos National Laboratory, \\ Thermonuclear Applications
Group (X-2), \\PO Box 1663, MS T085, \\Los Alamos, NM 87545 USA \\email: {\tt
joy@lanl.gov}}

\pubyear{2008}
\volume{250}  
\jname{Massive Stars as Cosmic Engines}
\editors{F. Bresolin, P.A. Crowther, \& J. Puls, eds.}
\begin{document}

\maketitle

\begin{abstract}
Luminous blue variables (LBVs) are characterized by semi-periodic episodes of
enhanced mass-loss, or outburst.  The cause of these outbursts has thus far been a mystery.  One
explanation is that they are initiated by $\kappa$-effect pulsations in the atmosphere
caused by an increase in luminosity at temperatures near the so-called ``iron
bump'' ($T \sim $ 200,000 K), where the Fe opacity suddenly increases.  Due to a
lag in the onset of convection, this luminosity 
can build until it exceeds the Eddington limit locally, seeding pulsations and possibly driving some
mass from the star.  We present some preliminary results from a parameter study
focusing on the conditions necessary to trigger normal S-Dor type (as opposed to
extreme $\eta$-Car type) outbursts.  We find that as $Y$ increases or $Z$
decreases, the pulsational amplitude decreases and outburst-like behavior,
indicated by a large, sudden increase in photospheric velocity, becomes likes
likely.

\end{abstract}

\firstsection

\section{Introduction}
Luminous blue variables (LBV) represent a short-lived phase
of massive star evolution characterized by episodes of intense mass loss, or
outburst.  There
are two main types of outburst behavior.  The rare eruptions such as
happened with $\eta$-Car and P-Cyg, occur every few decades to centuries.  They
last as long as a few tens of years and blow off $\geq 1 M_\odot$ of material
in a single episode.  The star also experiences an increase in luminosity during
this eruption phase.

In contrast, the more normal S-Dor-type episodes occur every few months to few
years.  They eject $\sim 10^{-4} M_\odot$ per event.  The LBV does {\it not}
become more luminous during these episodes.  For the purposes of this paper, we
will call the more extreme $\eta$-Car-like episodes eruptions and the less
extreme S-Dor-like episodes outbursts. 

\cite[Vink \& de Koter (2002)]{vinkdekoter02} determined that the outburst phase results from 
changes in the dominant ionization states of Fe in the wind which result from
changes in the stellar radius and 
effective temperature, but what causes the changes in the radius and temperature is
not clear.  In this paper we explore the hypothesis that it is due to
seeding from $\kappa$-effect pulsations arising from a large increase in luminosity due to
the increase in Fe opacity from the so-called ``hot iron bump'' at temperatures
near 200,000 K.  We will also discuss the 
possiblity that a super-Eddington luminosity in regions of the star near these
iron bump temperatures could directly drive mass from the star.

\section{Models}
We start with massive main sequence models in the mass range $50 M_\odot \leq
M_* \leq 80 M_\odot$.  We evolve the stars using the Iben evolution code
\cite[(Iben 1965)]{iben65}.  Important additions to the code are the Swenson SRIFF analytical
EOS, OPAL opacities using the Grevesse and Noels 93 mixture (as used in
\cite[Iglesias \& Rogers (1996)]{iglesiasrogers96}), and the empirical mass-loss
prescription of \cite[Nieuwenhuijzen \& de Jager (1990)]{ndj90}.  We vary the
mass-loss rate to allow different surface abundances 
at the end of the evolution.  More details of this and the other codes
used in this analysis are found in \cite[Guzik, Watson, \& Cox
(2005)]{guziketal05}.

Once we have a model with properties similar to an LBV star, we run the model 
through a pulsation analysis code. This code determines the first three harmonics of the 
$\kappa$-effect pulsations. The most unstable mode -- that is, the one with the largest growth 
rate, as determined by the fractional energy gain per period -- is then chosen for further analysis. The modes chosen for this analysis are shown 
in Table \ref{modetable}. We analyze this unstable mode using the 1D Lagrangian hydrodynamics 
code Dynstar \cite[(Cox \& Ostlie 1993)]{coxostlie93}. This code includes the time-dependent convection 
treatment formulated by \cite[Ostlie (1990)]{cox90} and described below.

\begin{table}
\begin{center}
\caption{Properties of analyzed modes}
\label{modetable}
{\scriptsize
\begin{tabular}{|c|c|c|c|}\hline
{\bf Y} & {\bf Z} & {\bf Mode$^1$} & {\bf Fractional Energy }\\
   &      &    & {\bf Gain / Period} \\ \hline
29 & 0.01 & 1H & 1.11 \\ \hline
29 & 0.02 & 1H & 2.07 \\ \hline
49 & 0.01 & 1H & 0.552 \\ \hline
49 & 0.02 & 1H & 2.22 \\ \hline
\end{tabular}
}
\end{center}
\end{table}

The models have log $L / L_\odot \approx 6.1$ and $T_{eff} = 1.6\times 10^4$ K.  We
vary the metallicity $Z$ and the helium abundance $Y$.  Since the pulsations are
driven by iron opacity, they are $Z$-dependent, with lower $Z$
resulting in smaller amplitudes.  As $Y$ is increased, the star effectively
becomes older and closer to the stable Wolf-Rayet phase,  also resultin
in smaller pulsation amplitudes.

\section{Time-Dependent Convection}
Our time-dependent convection model is described in detail in \cite[Ostlie
(1990)]{ostlie90}.  Here are the main highlights.

The time-dependent convection model is built upon standard mixing length theory
\cite[(Bohm {\bf ??}; Cox \& Giuli 1968)]{bohmyy,coxgiuli68}.  In the standard
theory the convective luminosity is 
\begin{equation}
\label{lceqn}
L_c = 4\pi r^2 \frac{4T\rho C_P}{g\ell Q} v_c^{0\;3},
\end{equation}
where $r, T, C_P, \rho,$ and $g$ are the local radius, temperature, specific
heat, density, and gravitational acceleration, respectively.  The mixing length
$\ell$ is assumed to be equal to the pressure scale height.  The parameter $Q$
defined
\begin{equation}
\label{qdef}
Q \equiv -\frac{\partial \log \rho}{\partial \log T} \vert_P .
\end{equation}
$Q = 1$ for an ideal gas with a constant mean molecular weight.  The
convective eddy velocity in the static case is
\begin{equation}
\label{vcstaticdef}
v_c^0 = \frac{1}{4}\sqrt{\frac{gQ\ell^2}{2H_P} f},
\end{equation}
where $H_P$ is the pressure scale height,
\begin{equation}
\label{fdef}
f \equiv [\sqrt{1 + 4A^2(\Delta - \Delta_{ad})} - 1] / A,
\end{equation}
\begin{equation}
\label{deltadef}
\Delta \equiv \frac{d \log T}{d \log P},
\end{equation}
\begin{equation}
A \equiv \frac{Q^{1/2} \: C_P \kappa g \rho^{5/2} \: \ell^2}{12 \sqrt{2} a c P^{1/2}
\: T^3},
\end{equation}
$\Delta - \Delta_{ad}$ is the superadiabatic gradient, and $\kappa$ is the local
opacity.

The static case, however, does not account for the effect of convective lag due
to the inertia of the material.  This lag is included in our model using a
quadratic Lagrange interpolation polynomial of $v_c$ over the current and
previous two time steps.

For zone $i$ and time step $n$ the time-dependent value of $v_c$ is
\begin{eqnarray}
\label{tdvcdef}
v_{c,\:i}^n & \equiv & \frac{(t' - t_{n-1})(t' - t_{n-2})}{(t_n - t_{n-1})(t_n -
t_{n-2})} v_{c,\:i}^0 + \frac{(t' - t_n)(t' - t_{n-2})}{(t_{n-1} - t_n)(t_{n-1}
- t_{n-2})} v_{c,\;i}^{n-1} \\
& + & \frac{(t' - t_n)(t'-t_{n-1})}{(t_{n-2}-t_n)(t_{n-2}-t_{n-1})}
v_{c,\:i}^{n-2}, \nonumber
\end{eqnarray}
where $v_{c\;i}^0$ is the value determined from equation \ref{vcstaticdef},
\begin{equation}
\label{tprimedef}
t' \equiv t_{n-1} + \tau(t_n - t_{n-1})
\end{equation}
and
\begin{equation}
\label{taudef}
\tau \equiv \frac{v_{c\;i}^n (t_n - t_{n-1})}{\ell} l_{fac}.
\end{equation}
The lag factor $l_{fac} \equiv 1$ in our models.

\section{Results}
\begin{figure}[ht]
\begin{center}
\includegraphics[width=3.4in]{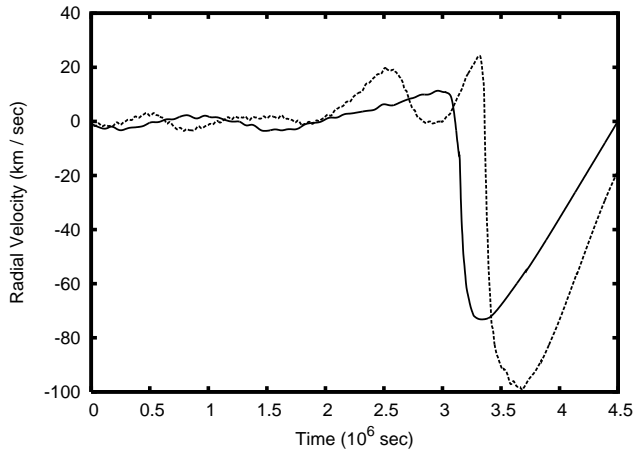}
\caption{Photospheric velocity as a function of time.  Solid line is $Z$ = 0.02,
dashed line is $Z$ = 0.01.  Negative velocity is outward.}
\label{metalfig}
\end{center}
\end{figure}

\begin{figure}[ht]
\begin{center}
\includegraphics[width=3.4in]{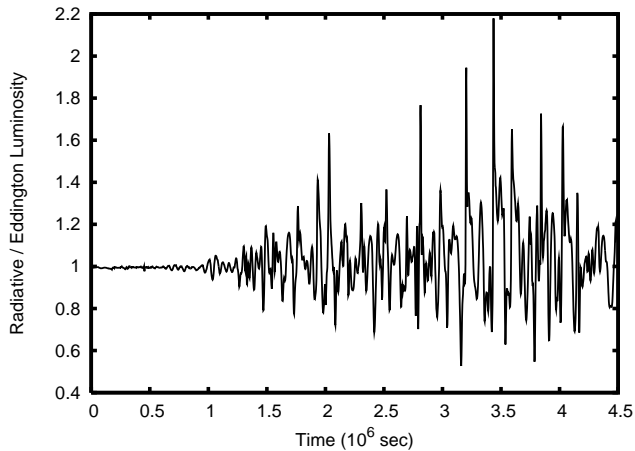}
\caption{Radiative luminosity at T $\sim$ 210,000 K relative to the Eddington
luminosity.}
\label{suped1}
\end{center}
\end{figure}

Figure \ref{metalfig} shows how the photospheric velocity $v_{phot}$ is affected
by changing the metallicity.  The figure
shows the photospheric velocity as a function of time for a model at solar
metallicity (defined here as $Z = 0.02$) and at half-solar, or LMC-like $Z$.  Both models exhibit
``outburst''-like behavior, in that extreme super-Eddington zones deep in the star
near 200,000 K (see Figure \ref{suped1}) result in a large, sudden increase in
the surface velocity.  As expected the model with the larger metallicity
experienced a larger velocity amplitude.

\begin{figure}[ht]
\begin{center}
\includegraphics[width=3.4in]{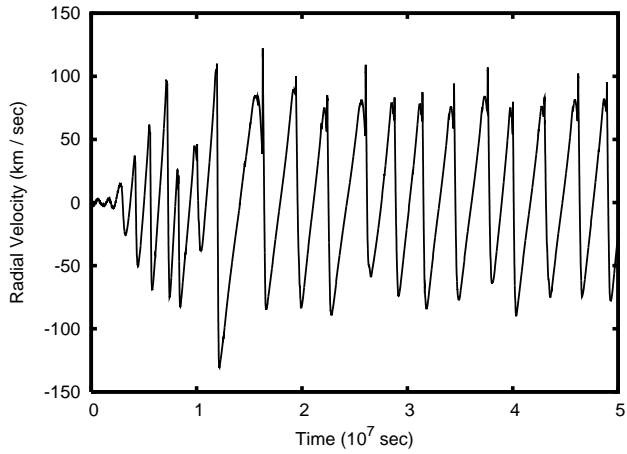}
\caption{Photospheric velocity as a function of time for the case $Y = 0.49, Z =
0.01$.}
\label{y49fig}
\end{center}
\end{figure}

\begin{figure}[ht]
\begin{center}
\includegraphics[width=3.4in]{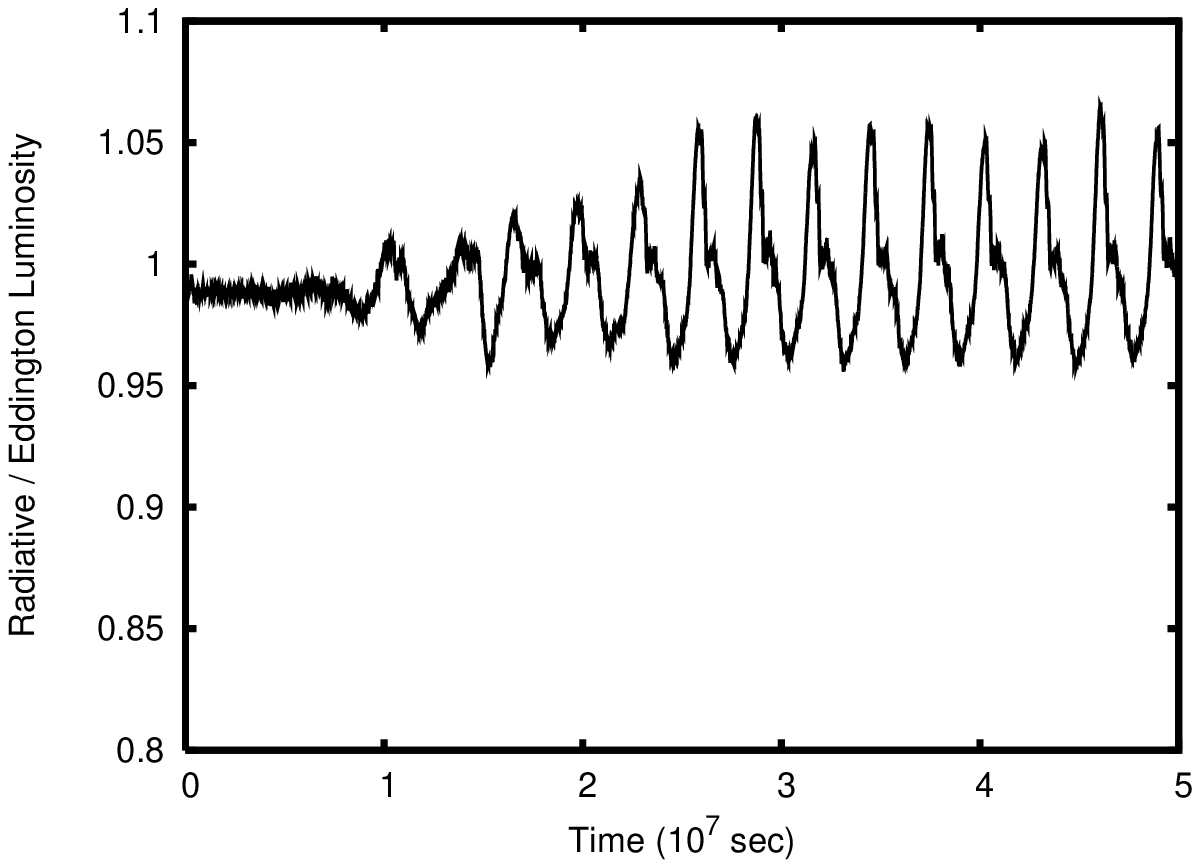}
\caption{Radiative luminosity at $T \sim$ 210,000 K relative to Eddington
for $Y = 0.49, Z = 0.01$.}
\label{suped2}
\end{center}
\end{figure}

The $Z = 0.01$ model with $Y$ increased to 0.49 is shown in Figure \ref{y49fig}.
In contrast to the ``outburst'' models, this model settles into steady
pulsations, and 
Figure \ref{suped2} shows that this model only becomes
super-Eddington by a few percent. Thus the star seems able to recover and settle
into steady, albeit very large amplitude ($v_{phot} = 100$ km s$^-1$) pulsations.

\section{Conclusions and Future Work}
We have preliminary results from a study on $\kappa$-effect pulsations in LBV
stars, with possible implications for S-Dor-like mass loss.  We have found that
regions in the star near $T \sim$ 200,000 K experience large super-Eddington
luminosities due to the increase of radiation from iron bump opacities.  We have
found that this can cause large amplitude pulsations at the surface.  At lower
$Y$ and higher $Z$, a large sudden increase in radial velocity is seen at the
surface.  Whether the pulsations or the sudden velocity increases evolve into
S-Dor-like outbursts is not clear.

The results presented here are still very preliminary.  Much more work is needed
to confirm our results.  Not the least of which is a zoning study.  We have used
60-zone models to improve the throughput, but it is not clear that this is
sufficient.  Runs at 120 zones and higher should be run in order to confirm that
the results are robust.

In addition it is not clear that our ``outbursts'' result in much, if any mass
loss directly.  As Owocki, et al. showed in these proceedings, the most likely
scenario for mass loss in a super-Eddington atmosphere is one in which the
material separates into clumps with the clumps essentially stationary and some
small amount of material driven above the escape speed between these clumps.  As
we have only a 1D analysis thus far, it is impossible for us to determine what
the clump size distribution would be.

This is the beginning of a more formal parameter study on the nature of these
pulsations and their connections, if any, to S-Dor behavior.  In addition to
more coverage of the parameter space, we eventually plan to extend the
hydrodynamic analysis to 2D to determine whether the super-Eddington nature of
the pulsations leads to enhanced mass loss through pores in the stellar
atmosphere.

\acknowledgements A. Onifer would like to that the IAU for partial support for
this presentation.

\begin{discussion}
\discuss{Langer}{How many pulsation cycles does it take to lose the pulsating
layer in a stationary wind, assuming standard radiation-driven wind mass-loss
rates?}

\discuss{Onifer}{The pulsation cycle for the model I showed was about 1 month,
and the Fe bump region is $\sim 10^{-3} M_\odot$ below the surface, so it would
take about 100 years or $\sim $ 1000 pulsation cycles to strip away the layer.}

\discuss{Humphreys}{Thanks for making the distinction between the normal or
classical LBVs like S Dor \& AG Car and the $\eta$ Car-like giant eruptions.
$\eta$ Car \& its rare relatives actually increase their luminosity during their
giant eruptions.  The classical LBVs maintain constant luminosity during their
optically thick wind stage.  Owocki was talking about $\eta$-Car-like eruptions.
It may only be a difference of degree, and the mechanism may be the same, but
what we observe is very different.  It is important to make the distinction.}

\discuss{Onifer}{Thank you, this is a very good point.}

\discuss{Hirschi}{If pulsations are driven by iron opacity, pulsations would be
strongly metallicity dependent.  Are there other sources of opacity at very low
metallicities to drive pulsations.}

\discuss{Onifer}{I'm not aware of any sources of opacity at similar temperatures
that could take over for iron at low metallicity.}
\end{discussion}


\begin{thebibliography}{}

\bibitem[Cox \& Giuli(1968)]{coxgiuli68}{Cox, A.N., Giuli, R.T.} 1968,
\textit{Principles of stellar structure}, (New York: Gordon and Breach)
\bibitem[Cox \& Ostlie (1993)]{coxostlie93}{Cox, A.N., Ostlie, D.A.} 1993,
\textit{Ap\&SS}, 210, 311
\bibitem[Guzik \etal (2005)]{guziketal05}{Guzik, J.A., Watson, L.S., Cox, A.N.}
2005, \textit{ApJ}, 627, 1049
\bibitem[Iben (1965)]{iben65}{Iben, I.J.} 1965, \textit{ApJ}, 142, 1447
\bibitem[Iglesias \& Rogers (1996)]{iglesiasrogers96}{Iglesias, C.A., Rogers, F.J.} 1996, \textit{ApJ},
464, 943
\bibitem[Nieuwenhuijzen \& de Jager (1990)]{ndj90}{Nieuwenhuijzen, H., de Jager,
C.} 1990, \textit{A\&A}, 231, 134
\bibitem[Ostlie (1990)]{ostlie90}{Ostlie, D.A.} 1990, in: J.R. Buchler (ed.),
\textit{Numerical Modeling of Nonlinear Stellar Pulsations: Problems and
Prospects}, (Dordect, Boston: Kluwer Academic Publishers) p. 89
\bibitem[Vink \& de Koter (2002)]{vinkdekoter02}{Vink, J.S., de Koter, A.} 2002, \textit{A\&A},
393, 543

\end{thebibliography}
\end{document}